\documentclass[a4paper,twocolumn,11pt]{quantumarticle}
\pdfoutput=1
\usepackage[utf8]{inputenc}
\usepackage[english]{babel}
\usepackage[T1]{fontenc}
\usepackage{amsmath}
\usepackage{hyperref} 

\usepackage{tikz}
\usepackage{lipsum}

\usepackage{braket}
\usepackage{graphicx}
\usepackage{siunitx}
\usepackage{dsfont}    
\usepackage{color}

\usepackage{indentfirst}
\usepackage{graphicx}
\usepackage{amssymb}
\usepackage{diagbox}

\usepackage{amsthm, mathtools}

\newcommand{\RN}[1]{%
	\textup{\uppercase\expandafter{\romannumeral#1}}%
}
\usepackage{dcolumn}
\usepackage{bm}

\newcommand{\nn}{\nonumber}

\def\a{\alpha}
\def\b{\beta}
\def\g{\gamma}

\def\r{\rho}

\def\p{\psi}

\def\P{\Psi}

\def\<{\langle}
\def\>{\rangle}

\def\ha{{\hat{a}}}
\def\had{{\hat{a}^\dagger}}

\def\hbd{{\hat{b}^\dagger}}

\usepackage{lineno}

\begin{document}

\title{Entangling three identical particles via spatial overlap}

\author{Donghwa Lee}
\affiliation{Center for Quantum Information, Korea Institute of Science and Technology (KIST), Seoul, 02792, Republic of Korea}
\affiliation{Division of Nano \& Information Technology, KIST School, Korea University of Science and Technology, Seoul 02792, Republic of Korea}
\author{Tanumoy Pramanik}
\affiliation{Center for Quantum Information, Korea Institute of Science and Technology (KIST), Seoul, 02792, Republic of Korea}
\author{Seongjin Hong}
\affiliation{Center for Quantum Information, Korea Institute of Science and Technology (KIST), Seoul, 02792, Republic of Korea}
\author{Young-Wook Cho}
\affiliation{Center for Quantum Information, Korea Institute of Science and Technology (KIST), Seoul, 02792, Republic of Korea}
\affiliation{Department of Physics, Yonsei University, Seoul 03722, Republic of Korea}
\author{Hyang-Tag Lim}
\affiliation{Center for Quantum Information, Korea Institute of Science and Technology (KIST), Seoul, 02792, Republic of Korea}
\affiliation{Division of Nano \& Information Technology, KIST School, Korea University of Science and Technology, Seoul 02792, Republic of Korea}
\author{Seungbeom Chin}
\thanks{Co-corresponding author}
\email{sbthesy@gmail.com}
\affiliation{Department of Electrical and Computer Engineering, Sungkyunkwan University, Suwon 16419, Korea}
\author{Yong-Su Kim}
\thanks{Corresponding author}
\email{yong-su.kim@kist.re.kr}
\affiliation{Center for Quantum Information, Korea Institute of Science and Technology (KIST), Seoul, 02792, Republic of Korea}
\affiliation{Division of Nano \& Information Technology, KIST School, Korea University of Science and Technology, Seoul 02792, Republic of Korea}
\maketitle

\begin{abstract}
Quantum correlations between identical particles are at the heart of quantum technologies. Several studies with two identical particles have shown that the spatial overlap and indistinguishability between the particles are  necessary for generating bipartite entanglement. On the other hand, researches on the extension to more than two-particle systems are limited by the practical difficulty to control multiple identical particles in laboratories. In this work, we propose schemes to generate two fundamental classes of genuine tripartite entanglement, i.e., GHZ and W classes, which are experimentally demonstrated with three identical photons. We also show that the tripartite entanglement class decays from the genuine entanglement to the full separability as the particles become more distinguishable from each other. Our results support the prediction that particle indistinguishability is a fundamental element for entangling identical particles.
\end{abstract}

\section{Intoduction}
Entanglement is a crucial resource of various quantum tasks such as quantum computation, quantum communication, and quantum sensing \cite{epr,monroe2002quantum,gross2007novel,vedral2014quantum,pirandola2015advances}. 
However, its authentic understanding and rigorous quantification still remain open issues \cite{Horodecki2009,regula2016entanglement}.
In particular, the entanglement phenomena become complicated when a quantum system  involves more than two parties~\cite{mikami2005new,walter2016multipartite,szalay2015multipartite,zhou2019detecting,nezami2020multipartite}, whereas the multipartite entanglement is beneficial in several aspects, e.g., nonlocality test \cite{Brunner2014}, multi-party quantum communication \cite{Hillery1999}, and quantum computation \cite{Ladd2010}.

Recently, the indistinguishability of quantum particles arouse strong interest as a useful resource for generating entanglement~\cite{tichy2013entanglement, killoran2014extracting, krenn2017entanglement, franco2018indistinguishability, barros2020entangling}, which was motivated by the debate on the rigorous quantification of entanglement in {\it identical particles}~\cite{balachandran2013entanglement, benatti2011entanglement, benatti2012bipartite, franco2016quantum, lourencco2019entanglement, chin2019entanglement}.
By the exchange symmetry, the total state of identical particles always seems to have a mathematical form of entanglement. Hence, we need rigorous criteria to discard the artificial entanglement from the physically relevant one in quantum systems. In the process of quantifying the physical entanglement in identical paticles, it has become clear that the particle indistinguishability is a necessary resource for the entanglement~\cite{franco2018indistinguishability,chin2019entanglement,barros2020entangling,sun2020experimental,chin2021taming}.


To emploit the indistinguishability for entangling identical particles, the spatial overlap between the particles is indispensable~\cite{paunkovic2004role,franco2018indistinguishability,chin2019entanglement}. The entanglement of identical particles should be observed at spatially distinguishable detectors, which must have nontrivial spatial coherence with the particles.  There are several researches on the quantitative relation of bipartite entanglement with particle indistinguishability and spatial coherence~\cite{franco2018indistinguishability,chin2019entanglement,tichy2013entanglement,barros2020entangling, sun2020experimental}.
Most notably, Ref.~\cite{barros2020entangling} verified theoretically and experimentally that the entanglement amount of two identical particles is a monotonically increasing function of indistinguishability and spatial coherence. 
There also exist some theoretical efforts for identical particle systems to generate the general $N$-partite entanglement~\cite{bellomo2017n,kim2020efficient,blasiak2019entangling,karczewski2019genuine,ju2019creating,castellini2019effects,chin2021graph}.
However, these researches are limited by the practical difficulty to control multiple identical particles in laboratories.

%


In this work, we push the limit by proposing schemes to generate genuine tripartite entanglement with identical particles, which is  experimentally demonstrated with three identical photons. Our schemes obtain two fundamental classes of tripartite genuinely entangled states, i.e., GHZ and W classes. Given that the GHZ and W classes are the only two genuine tripartite entanglement classes~\cite{dur2000}, our results show that \emph{all the genuine tripartite entangled states can be generated via spatial overlap among three identical particles.}

Furthermore, we investigate the quantitative relation of indistinguishability with the tripartite entanglement, which is accomplished by changing the relative indistinguishability of the particles and observing how the entanglement class varies accordingly. Our study shows that \emph{the tripartite entanglement class decays from the genuine entanglement to the full separability as the particles become more distinguishable.}
In particular, we find that three identical particles cannot carry the genuine entanglement if one of the particles is distinguishable from the others. Our result supports the prediction that the indistinguishability is an essential resource for the entanglement of identical particles.

\section{Theory}
Here,
we describe the general form of transformation operators for arbitrary $N$ particles that can produce entanglement. Then, we introduce specific transformations that generate two $N=3$ genuine entanglement classes, i.e., GHZ and W classes. We also analyze the relation of the tripartite entanglement class with indistinguishability.


We are interested in $N$ identical particles that can be distinguishable and have a two-level internal degrees of freedom. The state of a particle at  position $\p_i$ ($i\in \{1,2,\cdots, N\}$) is expressed as
\begin{align}\label{trans_special}
	|\P_i\>_{init.} = |\p_i,s_i,d_i\>
\end{align} where $s_i$ $(\in \{\uparrow,\downarrow\})$ and $d_i$ denote the internal state and distinguishability, respectively. To generate entanglement with identical particles, we need to spatially overlap the particles. We achieve the spatial overlap by transforming the initial state of the particles. We can generalize the transformation relation Eq.~\eqref{trans_special} so that the internal states can change according to the transformation:
\begin{align}\label{lqn_general}
	|\p_i,s_i,d_i\>=\sum_{j=1}^MT_{ij}|\phi_j,s_{ij},d_i\>, 
\end{align}
where $\phi_j$ denotes the position of $M$ detectors (modes),  $s_{ij}$ $(\in \{\uparrow,\downarrow\})$ denotes the internal state of a particle that leaves $\p_i$ for $\phi_j$, and the transformation amplitudes $T_{ij}$ are normalized as $\sum_{j}|T_{ij}|^2 =1$. $T_{ij}$ can be considered as the entries of an $(N\times M)$ transformation matrix $T$. In addition, we can construct an $(N\times M)$ internal state distribution matrix $S$ whose entries are given by $s_{ij}$. This type of entanglement generation for $N=M=2$ is studied in Refs.~\cite{franco2018indistinguishability,barros2020entangling}.


\begin{figure}[t]
	\centering
	\includegraphics[angle=0,width=3.5in]{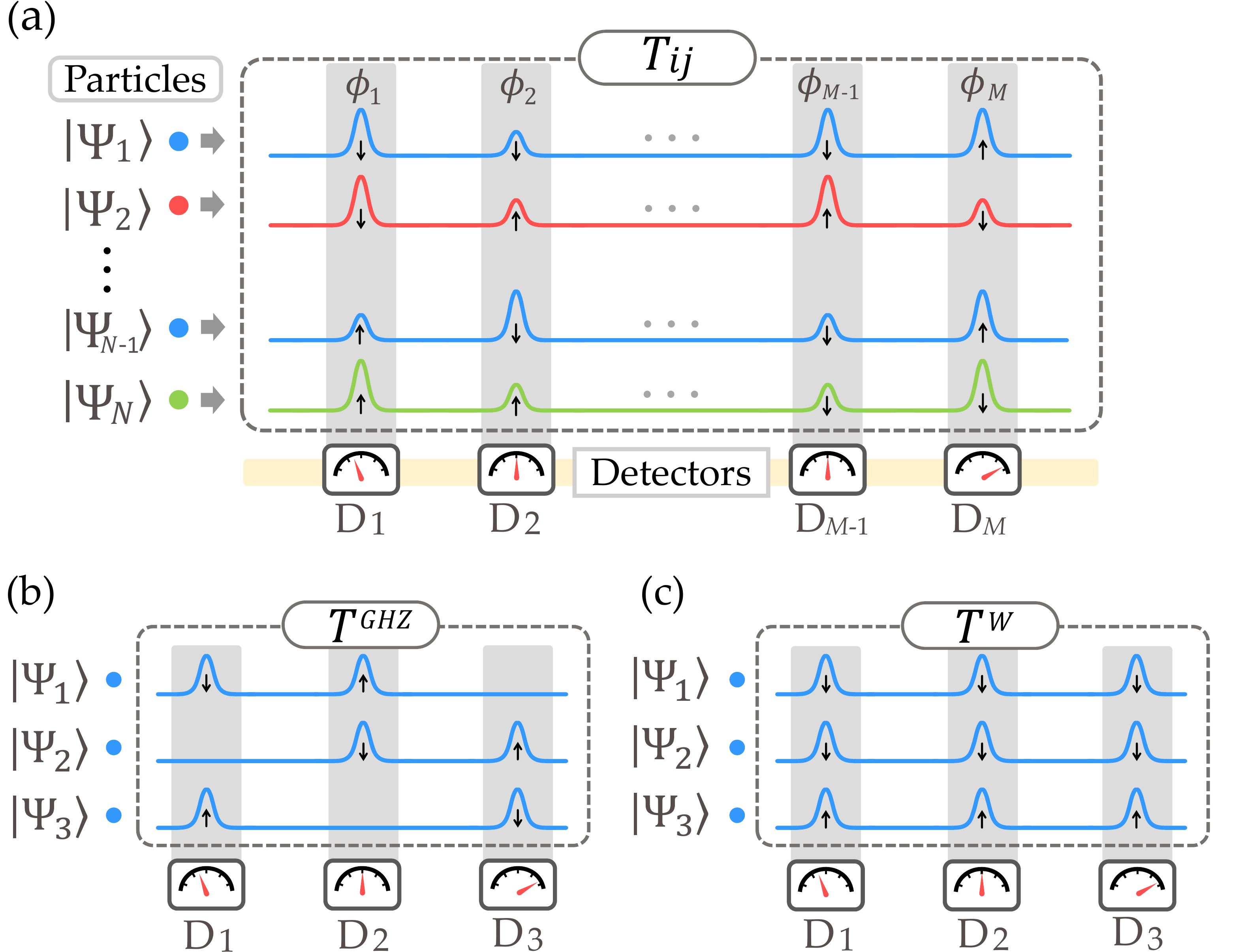}
	\caption{(a) The schematic diagram of a system that consists of N particles, M detectors and a transformation operator $T$. Each row (column) represents a particle (detector), and we restrict the particles to distribute in the detector region $\phi_{j}$ ($j\in \{1,2,\cdots M\}$) denoted as the gray zones. The wave amplitudes in the gray zones correspond to the spatial distribution amplitudes, and the internal states are given with arrows ($\downarrow$ and $\uparrow$). The distinguishability is depicted with different colors. (b) and (c) The transformation that generates GHZ class and W class entanglement ($T^{GHZ}$ and $T^W$), respectively. 
	}
	\label{diagram_main}
\end{figure}

To analyze the effect of $T$ and $S$ to the entanglement, 
we use the second quantization language. 
Then, Eq.~\eqref{lqn_general} is rewritten as
\begin{align}
	\had_{i}=\sum_{j=1}^M T_{ij}\hbd_{ij}, 
\end{align}
with the definitions $\had_{i}|vac\> = |\p_i,s_i,d_i\>$ and $\hbd_{ij}|vac\> = |\phi_j,s_{ij},d_i\>$. 
The $N$ particle transformation is expressed as
\begin{align}
	\prod^{N}_{i=1}\hat{a}^{\dagger}_{i}\vert vac \rangle=\prod^{N}_{i=1}\left(\sum^{M}_{j=1}T_{ij}\hat{b}^{\dagger}_{ij}\right)\vert vac \rangle.
	\label{equation_main}
\end{align} 
Figure \ref{diagram_main} (a) presents the schematic diagram of  $N$ identical particles that transform along $T$ and $S$ to reach $M$ detectors. Each particle is described as a wave that has non-zero amplitudes at the detector regions.

In the following, we focus on $N=M=3$ cases and postselect no particle-bunching states, i.e., states in which each detector observes one particle. We analyze the tripartite entanglement  according to the variations of the spatial overlap and particle distinguishability.
Particularly, we show that all the genuine tripartite entangled classes, i.e., GHZ and W states, can be generated in this setup. 



First, to generate the $N=3$ GHZ state, we consider the transformation relation Eq.~\eqref{equation_main} with
\begin{eqnarray}
	T^{GHZ}\equiv 
	\begin{pmatrix}
		\alpha_{1} & \alpha_{2} & 0 \\
		0 & \beta_{2} & \beta_{3} \\
		\gamma_{1} & 0 & \gamma_{3} \\
	\end{pmatrix},\nn 
	~~~~S^{GHZ}\equiv 
	\begin{pmatrix}
		\downarrow &\uparrow  & 0 \\
		0 & \downarrow  & \uparrow \\
		\uparrow & 0 & \downarrow
	\end{pmatrix},
	\label{matrix_GHZ}
\end{eqnarray}
\\

where $|\a_1|^2+|\a_2|^2=|\b_2|^2+|\b_3|^2=|\g_1|^2+|\g_3|^2=1$. The transformation is visualized in Fig.~\ref{diagram_main}~(b). 



After the postselection of no-bunching states, the unnormalized relevant state $|\P_{GHZ}\>$ is given by
\begin{align}
	\vert\Psi_{GHZ}\rangle =~ &\alpha_{1}\beta_{2}\gamma_{3}\vert\downarrow ,d_1\> |\downarrow ,d_2\> |\downarrow ,d_3\rangle\nn \\
	&+\alpha_{2}\beta_{3}\gamma_{1}|\uparrow ,d_3\> |\uparrow ,d_1\>|\uparrow ,d_2\>.
	\label{eq_ghz}
\end{align}
The detailed computation is provided in Supplemental Materials. We omit the spatial mode states and denote them with the state orders.
It is direct to see that 
$|\P_{GHZ}\>$ becomes a generalized GHZ state if all the particles are completely indistinguishable, i.e., $\langle d_i|d_j\rangle=1$ for $i,j\in\{1,2,3\}$.


To understand the effect of distinguishablity to $|\P_{GHZ}\>$, we compute the measurable density matrix $\r_{GHZ}$ that is obtained by tracing out the distinguishability terms of $|\P_{GHZ}\>$. Following the same method in Ref.~\cite{barros2020entangling}, we obtain
\begin{align}\label{rho_ghz}\nn
	\rho_{GHZ}=\\\nn
	\vert\alpha_{1}\beta_{2}&\gamma_{3}\vert^{2}\vert\downarrow\downarrow\downarrow\rangle\langle\downarrow\downarrow\downarrow\vert + \vert\alpha_{2}\beta_{3}\gamma_{1}\vert^{2}\vert\uparrow\uparrow\uparrow\rangle\langle\uparrow\uparrow\uparrow\vert\nonumber \\
	+~ \alpha_{1}&\beta_{2}\gamma_{3}\alpha^{*}_{2}\beta^{*}_{3}\gamma^{*}_{1}  \langle d_{3}d_{1}d_{2}\vert d_{1}d_{2}d_{3}\rangle \vert\downarrow\downarrow\downarrow\rangle\langle\uparrow\uparrow\uparrow\vert \nn \\
	+~ \alpha_{2}&\beta_{3}\gamma_{1}\alpha^{*}_{1}\beta^{*}_{2}\gamma^{*}_{3}  \langle d_{1}d_{2}d_{3}\vert d_{3}d_{1}d_{2}\rangle \vert\uparrow\uparrow\uparrow\rangle\langle\downarrow\downarrow\downarrow\vert.
\end{align} 
We can see that \emph{if one of the three particles is completely distinguishable, $\r_{GHZ}$ becomes fully separable.} For the complete definitions on the multipartite entanglement hierarchy from genuine entanglement to full separability, see  Refs.~\cite{dur2000classification,seevinck2008partial,szalay2012partial}. Indeed, by setting the third particle to be distinguishable ($\<d_1|d_3\> = \<d_2|d_3\>=0$) without loss of generality, we have $ \langle d_{3}d_{1}d_{2}\vert d_{1}d_{2}d_{3}\rangle=0$. Then, $\rho_{GHZ}$ becomes
\begin{align}
	\rho_{GHZ}\rightarrow
	\vert\alpha_{1}\beta_{2}\gamma_{3}\vert^{2}\vert\downarrow\downarrow\downarrow\rangle\langle\downarrow\downarrow\downarrow\vert + \vert\alpha_{2}\beta_{3}\gamma_{1}\vert^{2}\vert\uparrow\uparrow\uparrow\rangle\langle\uparrow\uparrow\uparrow\vert,
\end{align}
a fully separable state.



Second, we obtain the $N=3$ W class entanglement without transforming the internal states. 
We set the transformation operators are given by 
\begin{eqnarray}
	T^{W}=
	\begin{pmatrix}
		\alpha_{1} & \alpha_{2}  & \alpha_{3} \\
		\beta_{1} & \beta_{2}  & \beta_{3} \\
		\gamma_{1} & \gamma_{2}  & \gamma_{3}
	\end{pmatrix}, 
	~~~~S^{W}=
	\begin{pmatrix}
		\downarrow &\downarrow & \downarrow\\
		\downarrow & \downarrow & \downarrow \\
		\uparrow & \uparrow  & \uparrow
	\end{pmatrix}
	\label{matrix_W}
\end{eqnarray}
\\
where $\sum_{i=1}^3|\a_i|^2 =\sum_{i=1}^3|\b_i|^2 =\sum_{i=1}^3|\g_i|^2 =1$. The transformation is visualized in Fig.~\ref{diagram_main}~(c).



After the postselection, the unnormalized relevant state $|\Phi_W\>$ is given by
\begin{eqnarray}\nonumber
	\vert\Psi_{W}\> =&
	\a_1\b_2\g_3|\downarrow,d_1\>|\downarrow,d_2\>|\uparrow,d_3\>\nn \\ 
	&+ \a_1\b_3\g_2|\downarrow,d_1\>|\uparrow,d_3\>|\downarrow,d_2\>\nn \\
	&+\a_2\b_1\g_3|\downarrow,d_2\>|\downarrow,d_1\>|\uparrow,d_3\>\nn \\
	&+ \a_2\b_3\g_1|\uparrow,d_3\>|\downarrow,d_1\>|\downarrow,d_2\> \nn \\
	&+ \a_3\b_1\g_2|\downarrow,d_2\>|\uparrow,d_3\>|\downarrow,d_1\> \nn \\
	&+ \a_3\b_2\g_1|\uparrow,d_3\>|\downarrow,d_2\>|\downarrow,d_1\>.
	\label{eq_W}
\end{eqnarray} See Supplemental Materials for the detailed computation. We notice that 
$|\P_{W}\>$ becomes a generalized W state if the particles are completely indistinguishable.

To rigorously analyze the effect of distinguishability, we consider four different cases according to the relative distinguishablity of the three particles. We restrict our attention to the cases when the particles are completely indistinguishable or distinguishable with each other.

{\bf{Case~\RN{1}}} (All the particles are completely indistinguishable):  
In this case, Eq.~\eqref{eq_W} can be directly rewritten by omitting the distinguishabilty as
\begin{align}
	|\P^{\RN{1}}_W\> =~ &(\a_1\b_2\g_3+\a_2\b_1\g_3)|\downarrow\>|\downarrow\>|\uparrow\>\nn \\ 
	&+ (\a_1\b_3\g_2+ \a_3\b_1\g_2)|\downarrow\>|\uparrow\>|\downarrow\>\nn \\
	&+ (\a_2\b_3\g_1+ \a_3\b_2\g_1)|\uparrow\>|\downarrow\>|\downarrow\>, 
\end{align}  
which is a generalized W state with genuine tripartite entanglement.
{\bf Case~\RN{2}} (One of the two particles with internal state $|\downarrow\rangle$ i.e., $|\P_1\>$ or $|\P_2\>$, is distinguishable):
We can set the distinguishability as $(|d_1\>,|d_2\>,|d_3\>) = (|d_x\>,|d_y\>,|d_x\>)$ ($\<d_x|d_y\> =0$) without loss of generality. Then, we obtain the measurable density matrix $\r_{W}^{II}$ as
\begin{align}\nn
	\r_W^{\RN{2}} =~ &|\P^{\RN{2}}_{13|2}\>\<\P^{\RN{2}}_{13|2}| +|\P^{\RN{2}}_{12|3}\>\<\P^{\RN{2}}_{12|3}|\\
	&+|\P^{\RN{2}}_{23|1}\>\<\P^{\RN{2}}_{23|1}|,
	\label{W2}
\end{align}
where
\begin{align}
	&|\P^{\RN{2}}_{13|2}\>= \alpha_{1}\beta_{2}\gamma_{3}\vert\downarrow\downarrow\uparrow\rangle + \alpha_{3}\beta_{2}\gamma_{1}\vert\uparrow\downarrow\downarrow\rangle, \nn \\
	&|\P^{\RN{2}}_{12|3}\> =  \alpha_{1}\beta_{3}\gamma_{2}\vert\downarrow\uparrow\downarrow \rangle + \alpha_{2}\beta_{3}\gamma_{1}\vert\uparrow\downarrow\downarrow\rangle, \nn \\
	&|\P^{\RN{2}}_{23|1}\> =\alpha_{2}\beta_{1}\gamma_{3}\vert\downarrow\downarrow\uparrow \rangle+ \alpha_{3}\beta_{1}\gamma_{2}\vert\downarrow\uparrow\downarrow \rangle.
\end{align}
Since $\r_W^{\RN{2}}$ is a mixture of three bi-separable states $|\P^{\RN{2}}_{13|2}\>$,  $|\P^{\RN{2}}_{12|3}\>$, and  $|\P^{\RN{2}}_{23|1}\>$, it becomes a biseparable state. More specifically, it is in entanglement class 2.1 defined in  Ref.~\cite{seevinck2008partial} (bi-separable but not separable under any split of subsystems) and in class 2.1 defined in Ref~\cite{szalay2012partial} (bi-separable and mixed by all the three bipartitle subsystems).

\begin{table}[b]
	\setlength{\tabcolsep}{0.07in}
	\centering
	\begin{tabular}{|l|l|l|}
		\hline
		{\bf Case \RN{1}}          & \begin{tabular}{@{}c@{}}All three particles \\ indistinguishable. \end{tabular} & \begin{tabular}{@{}c@{}}~~~Genuinely \\ ~~~entangled \end{tabular} \\ 
		\hline
		{\bf Case \RN{2}} & \begin{tabular}{@{}c@{}} One particle in $\downarrow$ \\~~ distinguishable, \end{tabular} &  ~~Bi-separable \\
		\hline
		{\bf Case \RN{3}} & \begin{tabular}{@{}c@{}} The particle in $\uparrow$\\~~distinguishable. \end{tabular}  & Fully separable \\
		\hline
		{\bf Case \RN{4}} & \begin{tabular}{@{}c@{}} All three particles \\distinguishable. \end{tabular} & Fully separable \\[2ex]
		\hline
	\end{tabular}
	\caption{Four cases of particle distinguishability relations and the corresponding entanglement classes of the W class states. The state becomes more separable as the particles become more distinguishable.
	}
	\label{Case1to4}
	\end{table}

{\bf Case~\RN{3}} (The particle with internal state $|\uparrow\rangle$ is distinguishable) $\&$ {\bf Case~\RN{4}} (All the three particles are distinguishable with each other): 
The measurable density matrices of these two cases, $\r^{\RN{3},\RN{4}}_W$, are written as
\begin{align}
	\r^{\RN{3},\RN{4}}_W &= |\alpha_{1}\beta_{2}\gamma_{3}+\alpha_{2}\beta_{1}\gamma_{3}|^2\vert\downarrow\downarrow\uparrow\rangle\<\downarrow\downarrow\uparrow|\nn \\
	&  +|\alpha_{1}\beta_{3}\gamma_{2}+\alpha_{3}\beta_{1}\gamma_{2}|^2\vert\downarrow\uparrow\downarrow\rangle\<\downarrow\uparrow\downarrow| \nn \\
	& + |\alpha_{2}\beta_{3}\gamma_{1}+\alpha_{3}\beta_{2}\gamma_{1}|^2\vert\uparrow\downarrow\downarrow\rangle\< \uparrow\downarrow\downarrow|.
	\label{r_W^3}
\end{align}
This is a statistical mixture of $|\uparrow\uparrow\downarrow\rangle$, $|\uparrow\downarrow\uparrow\rangle$, and $|\downarrow\uparrow\uparrow\rangle$ states, hence fully separable. The detailed derivations of the W class entanglement are given in  Supplemental Materials.


We see that \emph{the entanglement class decays from genuine entanglement to full separability as the particles become more distinguishable with each other.} This tendency coincides with that of the bipartite case in Ref.~\cite{barros2020entangling}. {\bf Case \RN{2}} results in the most subtle form of bi-separable state, i.e., a mixture of all three bipartite subsystems ($12|3$, $13|2$, and $1|23$)~\cite{szalay2012partial}. And the fact that both {\bf Case \RN{3}} and {\bf \RN{4}} provide fully separable states shows that $|\P_W\>$ cannot be genuinely entangled if one particle becomes distinguishable.

\begin{figure*}[ht]
	\centering
	\includegraphics[angle=0,width=6.6in]{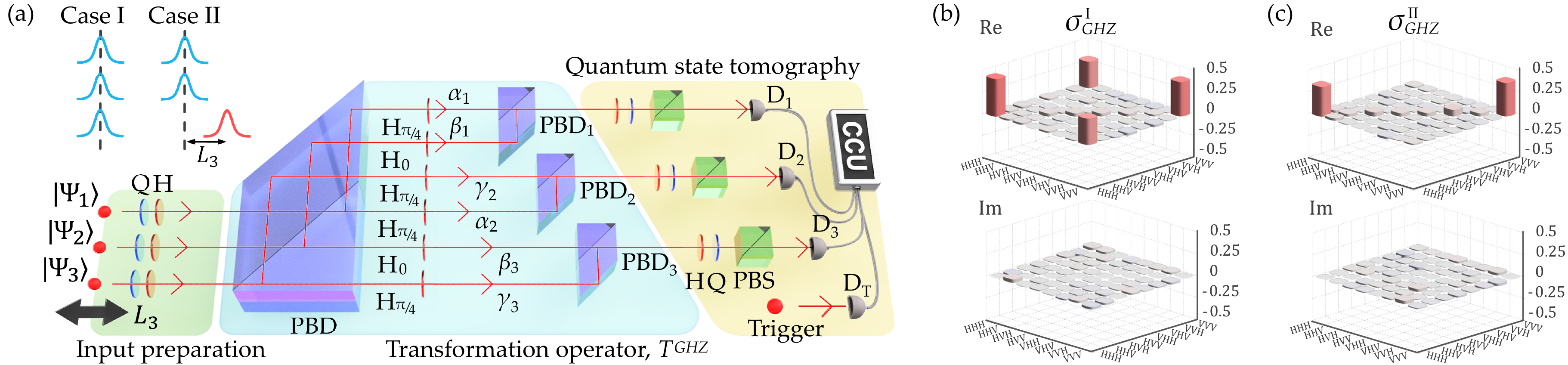}
	\caption{(a) The experimental setup to generate the GHZ class states. H: half waveplate, Q: quarter waveplate, BS: beamsplitter, PBD: polarization beam displacer, PBS: polarizing beamsplitter, APD: avalanche photo detector, CCU: coincidence counting unit. The experimentally reconstructed GHZ class sates. (b) $\sigma_{GHZ}^{\RN{1}}$ with $L_3=0$, and (c) $\sigma_{GHZ}^{\RN{2}}$ with $L_3\neq0$.}
	\label{ghz_setup_m}
\end{figure*}

\section{Experiment}
We have experimentally investigated entangling three identical particles using single-photon states from spontaneous parametric down-conversion. The horizontal and vertical polarization states, $|H\rangle$ and $|V\rangle$, serve as the spin states of $|\downarrow \rangle$ and $|\uparrow\rangle$, respectively. See the Supplemental Materials for details of preparation of initial three single-photon states for the following experiments.

Figure~\ref{ghz_setup_m} (a) presents the experimental setup to generate the GHZ class states using identical particles. The initial polarization states of the input photons $|s_1\rangle, |s_2\rangle$, and $|s_3\rangle$ are adjusted as $|D\rangle=\frac{1}{\sqrt{2}}\left(|H\rangle+|V\rangle\right)$ using sets of half- and quarter-waveplates (HWP and QWP). Then, the spatial wave function of the single photons are divided by a polarization beam displacer (PBD) which transmits and reflects the horizontal and vertical polarization states, respectively. After passing through HWPs at $\theta$ (${\rm H}_\theta$), the spatial wave functions are overlapped at PBDs such a way that the transformation relations of Eq.~(\ref{matrix_GHZ}) satisfy. Note that the outputs of ${\rm PBD}_j$ correspond to the spatial wave function $|\phi_j\rangle$. The polarization states of the three qubit particles were reconstructed by quantum state tomography (QST) using sets of HWP, QWP, and polarizing beamsplitters (PBS) in front of the single-photon detectors. The four-fold coincidence counts including a trigger photon are registered by a home-made coincidence counting unit (CCU)~\cite{park2015high, park2021arbitrary}. 
The particle distinguishability is adjusted by the temporal mode of the third photon which can be controlled by the relative delay $L_3$. As investigated in theory, we have considered two cases, i) all the photons have the identical temporal modes $L_3=0$, and ii) one particle is distinguished from the others, i.e., $L_3\neq0$. For the second case, we set $L_3=5~$mm which is much longer than the coherence length of the single photons. The experimentally reconstructed three-qubit states for all particles are indistinguishable and one particle is distinguishable are represented in Fig.~\ref{ghz_setup_m} (b) and (c), respectively. 
\begin{figure*}[ht]
	\includegraphics[angle=0,width=6.6in]{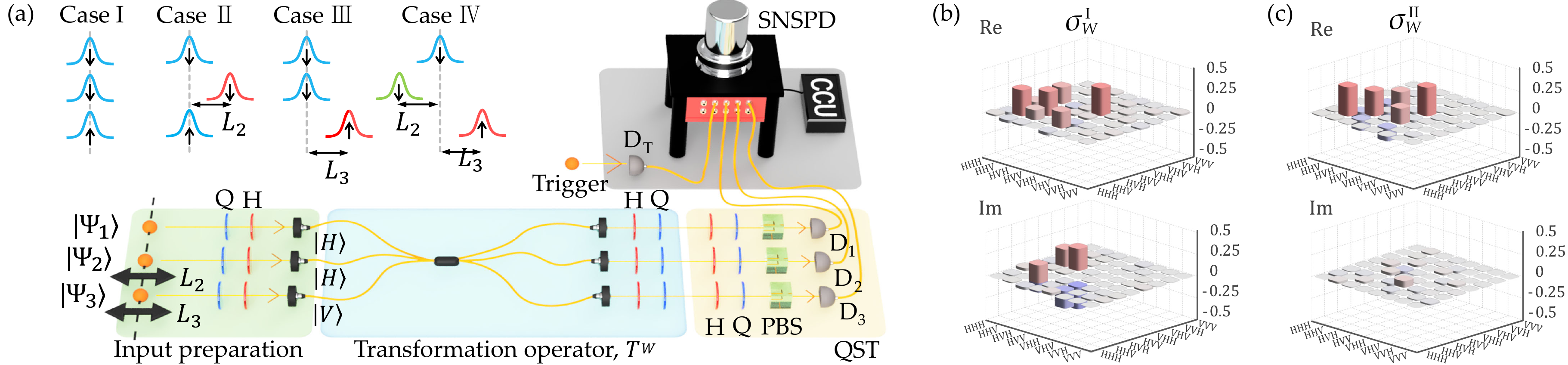}
	\centering
	\caption{(a) The experimental setup to generate W class states. H: half waveplate, Q: quarter waveplate, BS: beamsplitter, PBS: polarizing beamsplitter, SNSPD: superconducting nanowire single photon detector, CCU: coincidence counting unit. The experimentally reconstructed W class sates. (b) $\sigma_W^{\RN{1}}$ with $L_2=L_3=0$ and (c) $\sigma_W^{\RN{2}}$ with $L_2\neq0,~L_3=0$. See the Supplemental Materials for $\sigma_W^{\RN{3}}$ and $\sigma_W^{\RN{4}}$ } 
	\label{w_setup_m}
\end{figure*}
We now present the experimental setup to generate the W class states, see Fig.~\ref{w_setup_m}. The necessary transformation for the W state generation, Eq.~(\ref{matrix_W}) was achieved using an optical fiber based $3\times3$ symmetrical multiport (or a tritter) with two horizontally and one vertically polarized photon inputs. \cite{spagnolo2013three} 
The various {\bf Case \RN{1}}--{\bf \RN{4}} are implemented with the relative delay of the second and third photons, $L_2$ and $L_3$. For non-zero temporal delay, we set the values $|L_2|=|L_3|=5~$mm which is much larger than the coherence length of the single photon. At the outputs of the tritter, HWP and QWP compensate the polarization mode dispersion during the optical fiber transmission. Then, sets of HWP, QWP, and PBS are placed for three-qubit QST. The single-photons are detected by superconducting nanowire single-photon detectors (SNSPD). The experimentally reconstructed three-qubit density matrices are presented in in Fig. \ref{w_setup_m} (b), (c) and Appendix D.

\begin{figure}[h]
	\centering
	\includegraphics[angle=0,width=3in]{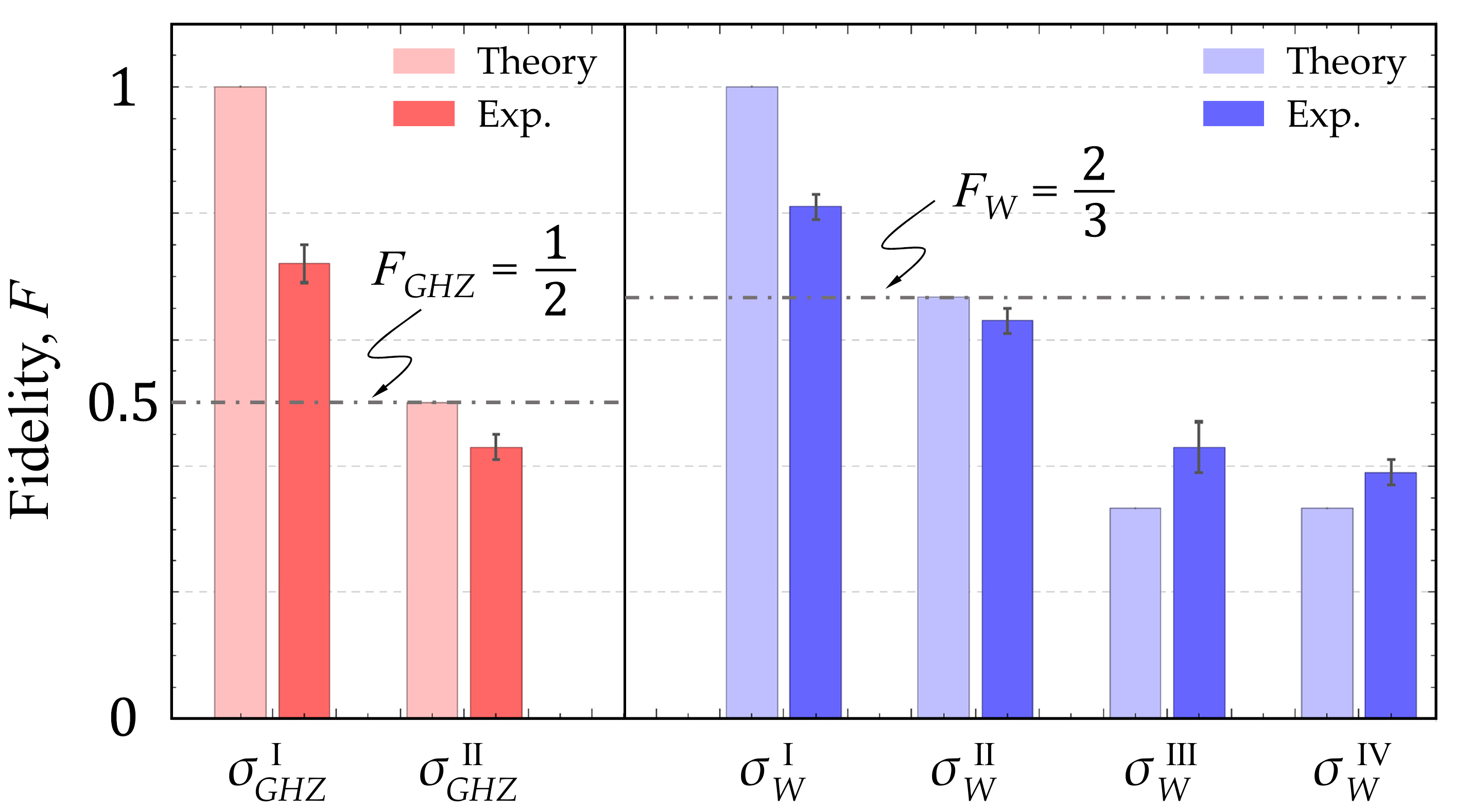}
	\caption{The theoretical and experimental fidelity values with ideal GHZ (red bars) and W states (blue bars). The horizontal lines of $F_{GHZ}=\frac{1}{2}$ and $F_W=\frac{2}{3}$ present the genuine tripartite entanglement bound of GHZ and W classes, respectively.}
	\label{graph_bar}
\end{figure}
Figure~\ref{graph_bar} shows the theoretical and experimental fidelity results between each case of $\sigma_{GHZ}~(\sigma_{W})$ and ideal GHZ (W) state with red (blue) color bars. The fidelity value of $F(\rho_{GHZ},\sigma_{GHZ})=\langle\Psi_{GHZ}|\sigma_{GHZ}^{\RN{1}}|\Psi_{GHZ}\rangle=0.72\pm0.03>0.5$ verifies that $\sigma_{GHZ}^{\RN{1}}$ possesses the genuine GHZ class three-partitie entanglement~\cite{Bourennane2004}. On the other hand, $\sigma_{GHZ}^{\RN{2}}$ resembles classical mixture of $|HHH\rangle$ and $|VVV\rangle$. The fidelity from the ideal GHZ state of $F(\rho_{GHZ},\sigma_{GHZ})=0.43\pm0.02<0.5$ does not show the genuine GHZ class tripartitie entanglement. The experimental results agrees well with our theoretical investigation.

In the W class, when all the photons are indistinguishable ({\bf Case \RN{1}}), the maximized fidelity between the generalized W state, $\vert\Psi_{W}^{\RN{1}}\rangle=\dfrac{1}{\sqrt{3}}(\vert HHV \rangle+e^{i\phi_1}\vert HVH \rangle+e^{i\phi_{2}}\vert VHH \rangle)$, and $\sigma^{\RN{1}}_W$ is $F(|\Psi_W^{\RN{1}}\rangle\langle\Psi_W^{\RN{1}}|,\sigma^{\RN{1}}_W)=0.81\pm0.02$ for $\phi_1=-0.21\pi$ and $\phi_2=0.28\pi$. Note that the non-zero relative phase $\phi_1$ and $\phi_2$ can be induced during the tritter channel transmission~\cite{weihs1996two,spagnolo2013three}. The genuine W class tripartite entanglement of $\sigma^{\RN{1}}_W$ can be witnessed by the fidelity with an ideal W state higher than $F=\frac{2}{3}$~\cite{Bourennane2004}. Note also that, in Fig. \ref{w_setup_m} (b), the reconstructed density matrix $\sigma^{\RN{1}}_W$ clearly shows non-zero off-diagonal elements which correspond to the coherence among $|HHV\rangle$, $|HVH\rangle$, and $|VHH\rangle$ states.
When one of two $|H\rangle$ photons is distinguishable with $L_2\neq0$ and $L_3=0$, the fidelity from an ideal W state of $F(|\Psi_W^{\RN{1}}\rangle\langle\Psi_W^{\RN{1}}|,\sigma^{\RN{2}}_W)=0.63\pm0.02<\frac{2}{3}$ shows that it does not possess the genuine W class three-party entanglement. However, its high fidelity with $\rho_W^{\RN{2}}$, $F(\rho_W^{\RN{2}},\sigma_W^{\RN{2}})=0.92\pm0.01$ presents that the experimentally reconstructed state $\sigma^{\RN{2}}_W$ is close to the bi-separable state $\rho_W^{\RN{2}}$. In Fig. \ref{w_setup_m} (c), it clearly shows smaller off-diagonal elements comparing to $\sigma^{\RN{1}}_W$ of the {\bf Case \RN{1}}.
As expected, both states of $\sigma^{\RN{3}}_W$ and $\sigma^{\RN{4}}_W$ have low fidelities with an ideal W state where $F(|\Psi_W^{\RN{1}}\rangle\langle\Psi_W^{\RN{1}}|,\sigma^{\RN{3}}_W)=0.43\pm0.04$ and $F(|\Psi_W^{\RN{1}}\rangle\langle\Psi_W^{\RN{1}}|,\sigma^{\RN{4}}_W)=0.39\pm0.02$ present that they have limited overlap with an ideal W state. On the other hand, they are closed to the classical mixture of $\rho_{W}^{mix}=\frac{1}{3}(|HHV\rangle\langle HHV|+|HVH\rangle\langle HVH|+|VHH\rangle\langle VHH|)$ as witnessed by high fidelities, $F(\rho_{W}^{mix},\sigma_W^{\RN{3}})=0.92\pm0.02$ and $F(\rho_{W}^{mix},\sigma_W^{\RN{4}})=0.94\pm0.01$, respectively.

\section{Conclusion}
We have provided schemes to construct two fundamental classes of tripartite genuinely entangled states, i.e., GHZ and W classes, and have experimentally implemented using identical photons. Our results verify that the particle indistinguishability plays an essential role for the entanglement of identical particles. We remark that the concepts of particle identity and spatial overlap forms fundamental aspects in quantum science, and thus, our photonic demonstration can be reproduced and extended by any quantum systems such as trapped atoms and ions~\cite{islam2015}, and solid state circuit quantum electrodynamics~\cite{gao2018}.

Our current work on the tripartite entanglement of identical particles can extend to general $N$ identical particle systems with more complicated spatial coherence. For instance, our entanglement generation schemes can be interpreted under the viewpoint of linear quantum networks (LQNs), which one can map into bipartite graphs~\cite{chin2021graph}. Therefore, we believe that our work initiates a new research avenue to generate multipartite entangled states with identical particles, which are essential for quantum information processing.

\section{Funding}
This work is supported by the National Research Foundation of Korea (2019M3E4A1079777, 2019R1A2C2006381 and 2019R1I1A1A01059964, 2021R1C1C1003625), MSIT/IITP (2020-0-00972 and 2020-0-00947), and a KIST research program (2E31531).

\section{Disclosures}
The authors declare no conflicts of interest.

\bibliographystyle{plain}

\onecolumn\newpage
\appendix

\section{The computations of tripartite no-bunching states}
\subsection{GHZ class}
By inserting the entries of $T^{GHZ}$ and $S^{GHZ}$ into Eq.~(4) of the main content, we obtain 

\begin{align}
	\ha_{1}^\dagger \ha_{2}^\dagger \ha_{3}^\dagger|vac\> &= 
	(\sum_{j=1}^3T^{GHZ}_{1j}\hbd_{1j})  (\sum_{k=1}^3T_{2k}^{GHZ}\hbd_{2k})  (\sum_{l=1}^3T_{3l}^{GHZ}\hbd_{3l})|vac\> \nn \\
	&=(\a_1\hbd_{11} + \a_2\hbd_{12})(\b_1\hbd_{22} + \b_2\hbd_{23})(\gamma_1\hbd_{31} + \gamma_2\hbd_{33})|vac\> \nn \\
	&=(\a_1|1,\downarrow,d_1\> + \a_2|2,\uparrow,d_1\>)(\b_1|2,\downarrow,d_2\> + \b_3|3,\uparrow,d_2\>)(\gamma_1|1,\uparrow,d_3\> + \gamma_3|3,\downarrow,d_3\>).
\end{align}
With the postselection of no-bunching states, the unnormalized relevant state $|\P_{GHZ}\>$ is given by
\begin{align}
	\vert\Psi_{GHZ}\rangle
	&= \a_1\b_2\gamma_3|1,\downarrow,d_1\>|2,\downarrow,d_2\>|3,\downarrow,d_3\>+ \a_2\b_3\gamma_1 |2,\uparrow,d_1\>|3,\uparrow,d_2\>|1,\uparrow,d_3\> \nn \\
	&=\alpha_{1}\beta_{2}\gamma_{3}\vert\downarrow d_1\> |\downarrow d_2\> |\downarrow d_3\rangle+\alpha_{2}\beta_{3}\gamma_{1}|\uparrow d_3\> |\uparrow d_1\>|\uparrow d_2\>,
	\label{eq_ghz}
\end{align} which is Eq.~(5) of the main content.
In the second equality of the above equation, we omit the spatial mode state and assume it to be denoted by the state order.


\subsection{W class}
The W class entanglement of identical particles can be obtained without transforming the internal states. 
We set $(s_1,s_2,s_3) = (\downarrow,\downarrow,\uparrow)$ and insert the  transformation operator in Eq. (6) of the main text into Eq. (4).
Then, the three-particle transformation relation is given by
\begin{align}
	\ha_{1}^\dagger \ha_{2}^\dagger \ha_{3}^\dagger|vac\> 	&= 
	(\sum_{j=1}^3T^{W}_{1j}\hbd_{1j})  (\sum_{k=1}^3T_{2k}^{W}\hbd_{2k})  (\sum_{l=1}^3T_{3l}^{W}\hbd_{3l})|vac\> \nn \\
	&=(\a_1\hbd_{11} + \a_2\hbd_{12}+\a_3\hbd_{13})(\b_1\hbd_{21} + \b_2\hbd_{22} + \b_3\hbd_{23}) (\gamma_1\hbd_{31} + \gamma_2\hbd_{32} +\g_3\hbd_{33})|vac\> \nn \\
	&=(\a_1|1,\downarrow,d_1\> + \a_2|2,\downarrow,d_1\> + \a_3|3,\downarrow,d_1\>) (\b_1|1,\downarrow,d_2\> + \b_2|2,\downarrow,d_2\> + \b_3|3,\downarrow,d_2\>)\nn\\
	&\quad \times (\gamma_1|1,\uparrow,d_3\> +\g_2|2,\uparrow,d_3\>+ \gamma_3|3,\uparrow,d_3\>).
\end{align}
After the postselection, the unnormalized relevant state is given by
\begin{eqnarray}\nonumber
	\vert\Psi_{W}\> =&
	\a_1\b_2\g_3|\downarrow,d_1\>|\downarrow,d_2\>|\uparrow,d_3\>+ \a_1\b_3\g_2|\downarrow,d_1\>|\uparrow,d_3\>|\downarrow,d_2\>\nn \\
	&+\a_2\b_1\g_3|\downarrow,d_2\>|\downarrow,d_1\>|\uparrow,d_3\>+ \a_2\b_3\g_1|\uparrow,d_3\>|\downarrow,d_1\>|\downarrow,d_2\> \nn \\
	&+ \a_3\b_1\g_2|\downarrow,d_2\>|\uparrow,d_3\>|\downarrow,d_1\> + \a_3\b_2\g_1|\uparrow,d_3\>|\downarrow,d_2\>|\downarrow,d_1\>,
	\label{eq_W}
\end{eqnarray} where the state orders denote the spatial modes. This is Eq.~(9) of the main content. 

\newpage

\section{The measurable density matrix of W states according to the distinguishability change}
In this section, we compute the density matrices of W class that can change according to the particle distinguishability, i.e., \textbf{Case~I $\sim$ IV}. 
Since \textbf{Case~I} is trivial to obtain, here we discuss the computations for  {\bf Case II, \RN{3}} and {\bf \RN{4}}.

{\bf Case~\RN{2}.}  We can set the distinguishability as $(|d_1\>,|d_2\>,|d_3\>) = (|d_x\>,|d_y\>,|d_x\>)$ ($\<d_x|d_y\> =0$) without loss of generality. Then, Eq.~\eqref{eq_W}  becomes
\begin{align}
	\label{sub_state_W_2}
	\vert\Psi^{\RN{2}}_{W}\rangle=&(\alpha_{1}\beta_{2}\gamma_{3}\vert\downarrow\downarrow\uparrow\rangle + \alpha_{3}\beta_{2}\gamma_{1}\vert\uparrow\downarrow\downarrow\rangle)\otimes\vert d_x d_y d_x \rangle
	+(\alpha_{1}\beta_{3}\gamma_{2}\vert\downarrow\uparrow\downarrow \rangle + \alpha_{2}\beta_{3}\gamma_{1}\vert\uparrow\downarrow\downarrow\rangle)\otimes\vert d_x d_x d_y \rangle \nn \\
	&+(\alpha_{2}\beta_{1}\gamma_{3}\vert\downarrow\downarrow\uparrow \rangle+ \alpha_{3}\beta_{1}\gamma_{2}\vert\downarrow\uparrow\downarrow \rangle)\otimes\vert d_yd_xd_x \rangle\\\nn
	\equiv & |\P^{\RN{2}}_{13|2}\>\otimes \vert d_x d_y d_x \rangle + |\P^{\RN{2}}_{12|3}\>\otimes \vert d_x d_x d_y \rangle+ |\P^{\RN{2}}_{23|1}\>\otimes \vert d_y d_x d_x \rangle. 
\end{align} Note that $|\P_{13|2}^{II}\> =\alpha_{1}\beta_{2}\gamma_{3}\vert\downarrow\downarrow\uparrow\rangle + \alpha_{3}\beta_{2}\gamma_{1}\vert\uparrow\downarrow\downarrow\rangle$ is a bi-separable pure state between subsystem 2 and the others, etc.
We obtain the measurable density matrix $\r^{\RN{2}}_W$, Eq.~(11) of the main content, by tracing out the distinguishability.

{\bf Case~\RN{3}.} The particle with internal state $|\uparrow\rangle$ is distinguishable, i.e., $(|d_1\>,|d_2\>,|d_3\>) = (|d_x\>,|d_x\>,|d_y\>)$.
Now Eq.~\eqref{eq_W} becomes
\begin{align}
	\label{sub_state_W_3}
	\vert\Psi^{\RN{3}}_{W}\rangle
	=& (\alpha_{1}\beta_{2}\gamma_{3}+\alpha_{2}\beta_{1}\gamma_{3})\vert\downarrow\downarrow\uparrow\rangle\otimes\vert d_xd_xd_y \rangle +(\alpha_{1}\beta_{3}\gamma_{2}+\alpha_{3}\beta_{1}\gamma_{2})\vert\downarrow\uparrow\downarrow\rangle\otimes\vert d_xd_yd_x\rangle \nonumber \\
	& +  (\alpha_{2}\beta_{3}\gamma_{1}+\alpha_{3}\beta_{2}\gamma_{1})\vert\uparrow\downarrow\downarrow\rangle\otimes\vert d_yd_xd_x \rangle
\end{align}  and   the measurable density matrix $\r^{\RN{3}}_W$ after tracing out the distinguishablity is given by
\begin{align}\label{r_W^3}
	\r^{\RN{3}}_W =& |\alpha_{1}\beta_{2}\gamma_{3}+\alpha_{2}\beta_{1}\gamma_{3}|^2\vert\downarrow\downarrow\uparrow\rangle\langle\downarrow\downarrow\uparrow|\\\nn 
	+&|\alpha_{1}\beta_{3}\gamma_{2}+\alpha_{3}\beta_{1}\gamma_{2}|^2\vert\downarrow\uparrow\downarrow\rangle\langle\downarrow\uparrow\downarrow| + |\alpha_{2}\beta_{3}\gamma_{1}+\alpha_{3}\beta_{2}\gamma_{1}|^2\vert\uparrow\downarrow\downarrow\rangle\langle \uparrow\downarrow\downarrow|.
\end{align}
We see that this state is a statistical mixture of $|\uparrow\uparrow\downarrow\rangle$, $|\uparrow\downarrow\uparrow\rangle$, and $|\downarrow\uparrow\uparrow\rangle$ states, and fully separable.
{\bf Case~\RN{4}.} All the three particles are distinguishable with each other, i.e., $(|d_1\>,|d_2\>,|d_3\>) = (|d_x\>,|d_y\>,|d_z\>)$ where $\<d_x|d_y\> = \<d_y|d_z\> =\<d_x|d_z\>  =0$.

Then, the pure state $|\P_{W}^{\RN{4}}\>$ and the measurable density matrix $\r^{\RN{4}}_W$ are given by
\begin{align}
	|\Psi^{\RN{4}}_{W}\rangle =& \alpha_{1}\beta_{2}\gamma_{3}\vert\downarrow\downarrow\uparrow\rangle\otimes\vert d_xd_yd_z\rangle + \alpha_{1}\beta_{3}\gamma_{2}\vert\downarrow\uparrow\downarrow\rangle\otimes\vert d_xd_zd_y\rangle + \alpha_{2}\beta_{1}\gamma_{3}\vert\downarrow\downarrow\uparrow\rangle\otimes\vert d_yd_xd_z\rangle \nn \\
	+& \alpha_{2}\beta_{3}\gamma_{1}\vert\uparrow\downarrow\downarrow\rangle\otimes\vert d_zd_xd_y\rangle+ \alpha_{3}\beta_{1}\gamma_{2}\vert\downarrow\uparrow\downarrow\rangle\otimes\vert d_yd_zd_x\rangle
	+ \alpha_{3}\beta_{2}\gamma_{1}\vert\uparrow\downarrow\downarrow\rangle\otimes\vert d_zd_yd_x\rangle.
\end{align}
and
\begin{align}
	\r^{\RN{4}}_W =& (|\alpha_{1}\beta_{2}\gamma_{3}|^2+|\alpha_{2}\beta_{1}\gamma_{3}|^2)\vert\downarrow\downarrow\uparrow\rangle\langle\downarrow\downarrow\uparrow| +(|\alpha_{1}\beta_{3}\gamma_{2}|^2\\\nn
	+& |\alpha_{3}\beta_{1}\gamma_{2}|^2)\vert\downarrow\uparrow\downarrow\rangle\langle\downarrow\uparrow\downarrow|  + (|\alpha_{2}\beta_{3}\gamma_{1}|^2+|\alpha_{3}\beta_{2}\gamma_{1}|^2)\vert\uparrow\downarrow\downarrow\rangle\langle \uparrow\downarrow\downarrow|.
	\label{r_W^4}
\end{align} 
Except for the postselection probability, $\rho_W^{\RN{4}}$ of Eq.~\eqref{r_W^4} is identical to $\rho_W^{\RN{3}}$ of Eq.~\eqref{r_W^3}, and thus a fully separable state. This results correspond to Eq. (13) of the main content.

\newpage

\section{The single photon sources for the experiments}

\noindent {\bf GHZ class experiments:} Four single-photon states at 780~nm are generated by two type-\RN{2} spontaneous parametric down-conversion (SPDC) setups using beta-barium borate (BBO) crystals. The intrinsic spectral and spatial indistinguishability of the photons are achieved by spectral and spatial mode filtering using 3~nm interference filters, and single-mode optical fibers, respectively. The indistinguishability of the photons is experimentally verified with the high Hong-Ou-Mandel interference visibility $V>0.85$. Then, one of four photons is used for trigger (T), and the others were sent to the experimental setup shown in Fig.~2 in the main text. 

\noindent {\bf W class experiments:} Four single-photon states at 1556~nm are prepared by SPDC using two 10~mm long type-II periodically-poled KTP crystals~\cite{lee2020reference}. After the spectral and spatial mode filtering with 3~nm bandpass filters and single-mode optical fibers, we achieved highly indistinguishable photons with the HOM interference visibility of $V > 0.9$ for all the photon pairs. Then, one of four photons is used for a trigger (T), and the others were sent to the experimental setup shown in Fig.~3 in the main text. 
\newpage
\section{The reconstructed density matrices of $\sigma^{\RN{3}}_W$ and $\sigma^{\RN{4}}_W$}

The experimentally reconstructed density matrices $\sigma_W^{\RN{3}}$ and $\sigma_W^{\RN{4}}$ for {\bf Case \RN{3}} and {\bf Case \RN{4}} are presented in Fig.~\ref{graph_bar}, respectively. As theoretically expected, both states have little amount of off-diagonal elements.

\begin{figure*}[h]
	\centering
	\includegraphics[angle=0,width=3.2in]{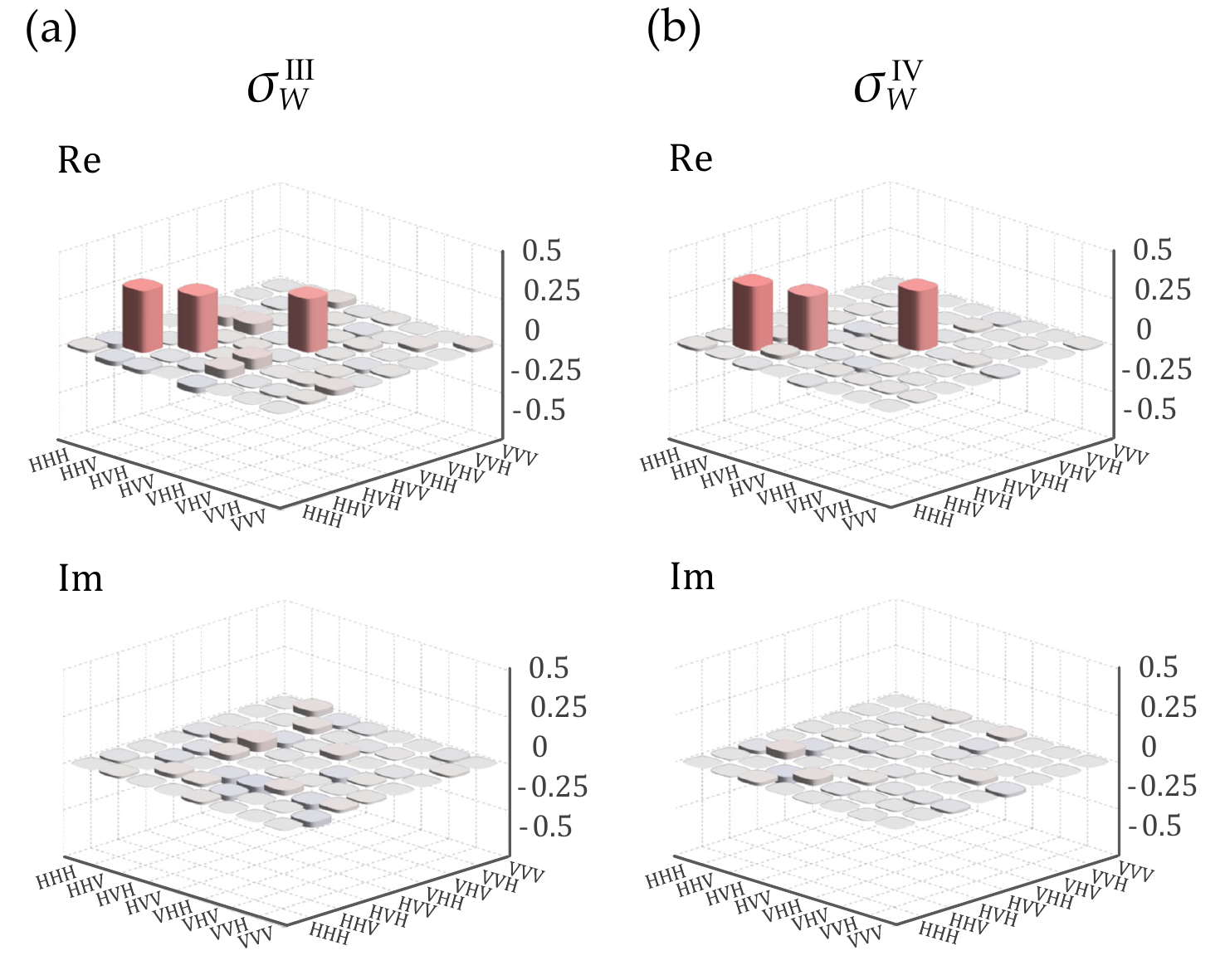}
	\caption{The experimentally reconstructed W class states. (a) $\sigma_W^{\RN{3}}$ with $L_2=0,~L_3\neq0$, and (b) $\sigma_W^{\RN{4}}$ with $L_2\neq0,~L_3\neq0$.}
	\label{M_exp_W3_4}
\end{figure*}

\end{document}